Quantum Mechanical Study of the Electronic Structure and Thermoelectric Properties of
Heusler Alloys

by

Deep Patel

A Thesis Presented in Partial Fulfillment
of the Requirements for the Degree
Master of Science

Approved November 2023 by the
Graduate Supervisory Committee:

Houlong Zhuang, Chair
Kiran Solanki
Huei-Ping Huang

ARIZONA STATE UNIVERSITY

December 2023

ABSTRACT

Heusler alloys were discovered in 1903, and materials with half-metallic characteristics have drawn more attention from researchers since the advances in semiconductor industry [1]. Heusler alloys have found application as spin-filters, tunnel junctions or giant magnetoresistance (GMR) devices in technological applications [1]. In this work, the electronic structures, phonon dispersion, thermal properties, and electrical conductivities of PdMnSn and six novel alloys (AuCrSn, AuMnGe, $Au_2MnSn$, $Cu_2NiGe$, $Pd_2NiGe$ and $Pt_2CoSn$) along with their magnetic moments are studied using ab initio calculations to understand the roots of half-metallicity in these alloys of Heusler family. From the phonon dispersion, the thermodynamic stability of the alloys in their respective phases is assessed. Phonon modes were also used to further understand the electrical transport in the crystals of these seven alloys. This study evaluates the relationship between materials' electrical conductivity and minority-spin bandgap in the band structure, and it provides suggestions for selecting constituent elements when designing new half-metallic Heusler alloys of $C1_b$ and $L2_1$ structures.

ACKNOWLEDGMENTS

I would like to thank  Arizona State University's Research Computing  for providing the necessary computing resources to carry out the ab initio calculations on Sol and Agave Clusters. I would also like to  thank Dr. Christopher Muhich for his advice  on the band structure calculations. I thank Dr. Houlong Zhuang, Dr. Kiran Solanki and Dr. Huei-Ping Huang for serving on my graduate committee.

I would like to extend my gratitude to my parents Kiritbhai Patel, Asmitaben Patel and my younger brother Kush Patel for their unwavering support in my research journey.

# TABLE OF CONTENTS

LIST OF TABLES

LIST OF FIGURES

v

CHAPTER 1

INTRODUCTION

Heusler alloys have been an intriguing subject of research due to their crystal structure and unique set of electronic, thermoelectric, and magnetic properties since their discovery in 1903 by a German mining engineer, Friedrich Heusler [2]. Friedrich Heusler's work in the early 20$^{th}$ century laid the foundation of research on the said group of materials, hence called Heusler alloys [2]. These alloys are distinguished in two distinct groups based on their crystal structure and constituent atoms. These two groups are Heusler alloys with $X_2YZ$ composition and Half-Heusler alloys with XYZ composition [1]. The alloys with $X_2YZ$ composition have been observed to have $L2_1$ type lattice structure which corresponds to Fm-3m space group (225 in the international tables of crystallography) [1,2]. The atomic configuration of both of these structures is shown in Figure 1 [1].

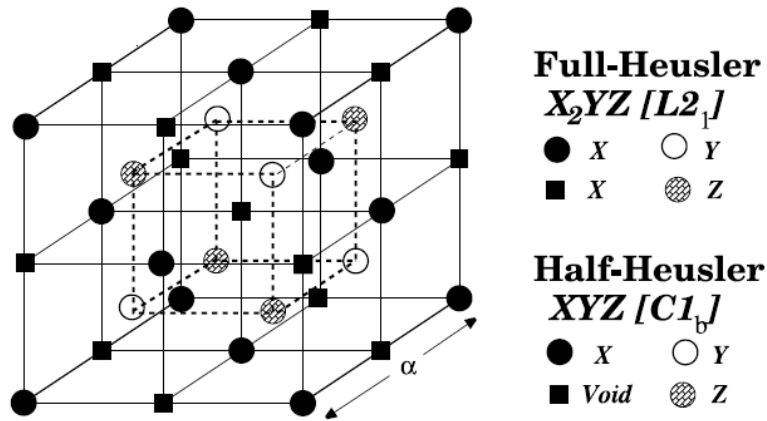

*Figure 1 Lattice Structure of Half-Heusler and Full-Heusler alloys.[1]*

The alloys with XYZ composition have been observed to have C1$_b$ type lattice structure, which corresponds to F43m space group (216 in the international tables of crystallography). XYZ alloys are also known as Half-Heusler Alloys (HHA) and Half Metallic Ferromagnets (HMF) [1,2]. X is a heavy transition element; Y is a rare earth element, or a transition metal and Z is a main group element [3]. Many of the alloys in the Heusler family are found to have 100% spin polarization of their valence electrons [1]. These materials showcase a set of electronic properties such that they are considered semi-metallic [2,3]. In these alloys, the band structure of majority spin showcases metallic behavior and that of minority spin showcases semiconductor-like characteristics [1,3]. In minority spin configuration, these materials have been observed to have a bandgap [1,3]. This phenomenon leads to lower electrical conductivity in the Heusler alloys relative to the other alloys which do not exhibit half-metallic properties. Heusler alloys have found application as spin-filters, tunnel junctions or giant magnetoresistance (GMR) devices in technological applications [1]. They are also used as thermoelectric materials due to their thermoelectric properties caused by the unique electron structure [4]. Due to such unconventional structural, electrical, and electronic properties, Heusler alloys continue to be studied for thermoelectric and spintronic applications [1,4].

## 1.1 Electronic Structure Theory

### 1.1.1 Many-Body Schrödinger's Equation

To describe a system of atoms or molecules made of n electrons and N nuclei, located at the coordinates {$r_1$, $r_2$, …, $r_n$} and {$R_1$, $R_2$, …, $R_N$} respectively, the time-Independent multi-body Schrödinger's equation is used in the ab initio approach [5]. The time-independent multi-body Schrödinger's equation is given as follows [5].

$$\hat{H}\Psi = E\Psi$$

Here, $\Psi$ is the wavefunction and it is the function of spatial coordinates of electrons and nuclei. $\hat{H}$ is the Hamiltonian operator and it is defined as follows [5,6].

$$\hat{H} = T^n + T^e + V^{e-e} + V^{n-n} + V^{e-n}$$

$$T^n = -\sum_{I=1}^{N} \frac{\hbar^2}{2M_I} \cdot \nabla_{R_I}^2$$

$$T^e = -\sum_{i=1}^{n} \frac{\hbar^2}{2m_i} \cdot \nabla_{R_i}^2$$

$$V^{e-e} = \frac{1}{4\pi\varepsilon_0} \sum_{i=1}^{n} \sum_{j>i} \frac{e^2}{r_{ij}}$$

$$V^{e-n} = \frac{-1}{4\pi\varepsilon_0} \sum_{i=1}^{n} \sum_{I=1}^{N} \frac{Z_I e^2}{|r_i - R_I|}$$

$$V^{n-n} = \frac{1}{4\pi\varepsilon_0} \sum_{I=1}^{N} \sum_{J>I} \frac{Z_I Z_J e^2}{R_{IJ}}$$

In the equations above, $T^n$ and $T^e$ terms represent the kinetic energy of the nuclei and electrons respectively [6]. $V^{e-e}$ term represents the potential energy due to Coulombic interaction between electrons, $V^{e-n}$ term represents the Coulombic potential due to interaction between electrons and nuclei and $V^{n-n}$ term represents the coulombic potential due to nuclei-nuclei interaction. $M_I$ and $Z_I$ are the mass and charge of the $I^{th}$ nucleus, e and m are electronic charge and electronic mass respectively. The many-body Schrödinger's equation represents high degree-of-freedom quantum mechanical interaction of the charged particles in the system of interest [5,6]. $\Psi$ is the wavefunction and E is energy of the system [5].

### 1.1.2 Density Functional Theory

With the introduction of Born-Oppenheimer approximation [5], the multibody Schrödinger's equation has been significantly simplified. For the solution of Schrödinger's equation, wavefunction methods scale with $N^4$ order or higher, which has extremely high computational cost. Density Functional Theory, which is based on the foundation laid by Hohenberg-Kohn theorems [5,6], states that the electronic ground state energy of the system is a functional of the electron density n(r) [5].

$$n(r) = \sum_{i=1}^{N_e} |\psi_i(r)|^2$$

$$E_0 = \sum_{i=1}^{N_e} \varepsilon_i - \frac{e^2}{2} \int \int \frac{n(r_1)n(r_2)}{|r_1 - r_2|} dr_1 \, dr_2 + E_{xc}[n(r)] - \int n(r)V_{xc}(r) \, dr$$

In the above equation, $E_0$ is the ground state energy of the system, $E_{xc}$ is the exchange & correlation energy and $V_{xc}$ is the exchange and correlation potential. There exists different exchange and correlation functionals such as Local Density Approximation (LDA), Generalized Gradient Approximation (GGA) such as "Becke, 3-parameter, Lee–Yang–Parr" (B3LYP) [7], Perdew-Burke-Ernzerhof (PBE) [8] etc. For further information on multi-body Schrodinger's equation and density functional theory, reader is referred to [5].

## 1.2 Electrical Conductivity

When materials are subjected to a unidirectional applied electric field, electrical transport takes place. In the effect of such nonequilibrium condition, the electric current density ($J$) and the electric field ($E$) are related as follows [9].

$$J = \sigma E$$

The proportionality constant $\sigma$ in above equation is electrical conductivity of the material.

Bloch waves are the wavefunctions/periodic potentials that provide the solution to the quantum mechanical problem of electronic ground state in a periodic structure (crystalline solids) [6]. In semi-classical approach a wave-packet made by superposition of Bloch waves can travel in the crystal infinitely and hence the crystal can ideally have infinite electrical conductivity [6]. However, in practice, lattice is anything but ideal. Crystal structures are prone to impurities, boundaries, dislocations, and other imperfections [6]. The interaction of the electron with other ions (electrons, nuclei and impurities) in the system introduces the electrical resistance [9].

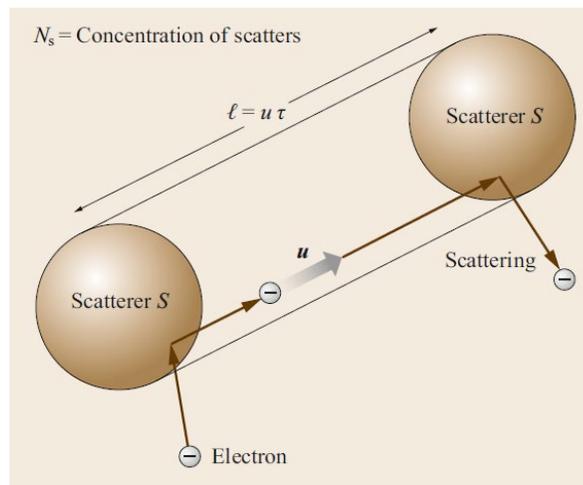

*Figure 2 Electron scattering in crystal lattice. [9]*

The electrons essentially have free motion between the scattering events with other ions. The average time between these scattering events is defined as relaxation time ($\tau$) [9]. The mean electron velocity between these scattering events is termed as thermal velocity ($v_{th}\ or\ u$) and the mean distance travelled by electron between consecutive scattering events is termed as mean free path ($l$) [9]. Figure 2 shows a schematic diagram of the thermal velocity ($u$) and the mean free path (l). Average thermal velocity of electron in semiconductors is given by $\sqrt{\frac{3k_BT}{m_e}} \approx 10^5$ m/s and in metallic materials, it is given by $\sqrt{\frac{2E_F}{m_e}} \approx 10^6$ m/s [9]. Electrons, when subject to external electric or magnetic field, acquire another velocity which is termed as drift velocity, and it is given by following equation [9]. The drift velocity is very small compared to thermal velocity of the electrons [9].

$$v_d = \frac{1}{N}\sum_{i=l}^{N} v_i$$

From above equation, drift velocity and current density are related by the equation $J = -env_d$ and $v_d = -\mu_d E$ [9], where $E$ is the applied electric field [9].

According to Matthiessen's rule, the average scattering time ($\tau$) can be related to $\tau_l$ – mean scattering time due to scattering with phonons and $\tau_I$ – mean scattering time due to scattering with impurities by following equation [9].

$$\frac{1}{\tau} = \frac{1}{\tau_l} + \frac{1}{\tau_I}$$

### 1.2.1 Boltzmann Transport Equation

In the Boltzmann's theory of electronic transport, the electronic system is described by a distribution function $f(k, r, t)$, such that the number of electrons in 6-dimensional volume $d^3k \, d^3r$ at time t is given by $\frac{1}{4\pi^3} f(k, r, t) d^3k \, d^3r$ [9]. The $k$ and $r$ are the special coordinates and $t$ is the time in the $f(k, r, t)$. The $f(k, r, t)$ is only dependent on energy and reduces to fermi distribution $f_0$, where the probability of occupation of states with momenta +k equals that for states with -k & $f_0(k)$ is symmetric about k=0. The Boltzmann Transport Equation is given as follows [9].

$$\frac{\partial f}{\partial t} = \left(\frac{\partial f}{\partial t}\right)_{diff} + \left(\frac{\partial f}{\partial t}\right)_{drift} + \left(\frac{\partial f}{\partial t}\right)_{scatt}$$

$$\frac{\partial f}{\partial t} = -v \cdot \nabla_r f - \dot{k} \cdot \nabla_k f + \left(\frac{\partial f}{\partial t}\right)_{scatt}$$

The $-v \cdot \nabla_r f$ term represents the diffusion through a volume element $d^3r$ about point $r$ in phase space due to $\nabla_r f$. The term $\dot{k} \cdot \nabla_k f$ represents drift through a volume element $d^3k$ about point $k$. More comprehensive information on BTE can be found in [9].

In 1929, the theory of electrons was presented by Bloch [7]. He also discussed the electron-lattice vibration interactions in the context of electron scattering, now known as electron-phonon (e-ph) interactions [9]. Based on Bloch's work on linearized BTE, electrical conductivity in the case of steady state current can be given by following equation [9].

$$\sigma = \frac{e^2}{3V} \sum_{nk} \left(-\frac{\partial f_{nk}^0}{\partial \varepsilon_{nk}}\right) v_{nk} \cdot v_{nk} \tau$$

In the equation above, $f_{nk}^0$ is fermi-dirac distribution function and $\tau$ is relaxation time. Before the ab initio methods became popular, relaxation time was assumed to be constant for simplification [9]. However, with the development of density functional perturbation theory (DFPT) [10] method to compute e-ph interactions, it became a standard practice to calculate state-resolved relaxation time from the equation of $\tau_{nk}^{-1}$ by equation 3.7 in Zhenkun Yuan's work on electrical conductivity [6].

## 1.3 Thermal Conductivity

Conduction of heat in a crystalline lattice is the phenomenon of transfer of kinetic energy through the scattering of electrons and atoms of lattice [9]. Hence from the ab initio perceptive, electron-phonon (e-ph) and phonon-phonon (ph-ph) interactions caused by thermal excitation of the lattice are crucial for thermal conduction. Thermal conductivity of a material is constituted by two components, one is electronic thermal conductivity ($\kappa_e$) and the other is lattice thermal conductivity ($\kappa_l$) [9].

# CHAPTER 2

## METHODOLOGY

The ab initio study of the Heusler and Half-Heusler alloys in this work includes electron structure, phonon dispersion, thermal properties, and electrical and thermal conductivities. The Vienna Ab initio Simulation Package (VASP 5.4.4) [11-14] was utilized to perform the first principles calculations. The Density Functional Theory (DFT) [5,8] with Perdew-Burke-Ernzerhof (PBE) functional [8] was chosen for the ab initio calculations. The kinetic energy cut-off of 600 eV was used for all calculations. The convergence criteria for total energy and ionic relaxation convergence were set to 1.0 x $10^{-6}$ eV and 1.0 x $10^{-2}$ eV/Å respectively. All calculations were spin-polarized as the concerned class of materials are expected to have 100% spin polarization of valence electrons.

For structural relaxation, a gamma-centered, automatically generated 10 x 10 x 10 k-point grid was used. For phonon calculations, a gamma-centered, automatically generated 2 x 2 x 2 k-point grid was used. For a 2 x 2 x 2 sized supercell of each alloy considered in current work. The finite difference method [15] was used to perform the phonon calculations as the density functional perturbation theory [10] takes an unperturbed supercell in input and performs the calculations by adding small vibrations to the atoms in all degrees of freedom in a single computational job, while the finite difference method [15] reduces the computational requirements of one single job as each degree of freedom of atomic vibration is computed in a separate computational tasks. For the Phonon calculations, the ionic convergence criteria of $10^{-3}$ eV/Å was chosen. For the electronic structure calculations, 100 points were taken on each high symmetry line of the lattice in reciprocal space to maintain the smoothness of the bands.

A python module vaspkit [16] was used to post-process the data of Partial and Total Density of States (PDOS and TDOS) and restructure the data of band structures from the results of self-consistent field (SCF) calculations carried out in VASP. Python module phonopy [17,18] was used to generate the displacements of atoms in supercell for the finite difference approach chosen for the phonon calculations. The force constants, specific heat, entropy Gibbs free energy and phonon dispersion spectra were calculated using phonopy [17,18].

For the calculation of electrical conductivity and electronic component of thermal conductivity, ab initio SCF calculations were carried out on the 2 x 2 x 2 supercells of the three $C1_b$ phase alloys and four $L2_1$ phase alloys with a k-point grid of 2 x 2 x 2 gamma-centered autogenerated mesh. Boltzmann Transport Phenomenon [6,9] was employed for the calculation of conductivities. Reader is advised to refer to [6,9] for a more comprehensive study of Boltzmann Transport Equation (BTE). BoltzTraP2 [19,20] python module was used to calculate thermal and electrical conductivities using BTE approach from the results of SCF calculations. It should be noted that the BoltzTraP2 module makes a primary assumption to simplify the process of calculation of thermoelectric properties, which is unit relaxation time ($\tau_0$) assumption. $\tau_0$ is an entity which is dependent on band index and k-vector [19,20]. However, many studies have proven the direction independence of the relaxation time [19,20]. Hence $\tau_0$ is considered constant (unity) for isotropic conditions to introduce further simplification [19,20].

# CHAPTER 3

## RESULTS AND DISCUSSION

### 3.1 Lattice Parameters

The lattice parameters of the relaxed geometry of Half-Heusler alloys crystallizing in $C1_b$ phase and Full-Heusler alloys crystallizing in $L2_1$ phase corresponding to F43m (216) and Fm-3m (225) space groups respectively are discussed in this section. The following tables 1 and 2 describe the lattice parameters of the seven alloys considered in this work.

*Table 1 Lattice Parameters of Half Heusler Alloys*

| $C1_b$ Alloy | Lattice Parameter (Å) |
|:---:|:---:|
| AuCrSn | 6.364 |
| AuMnGe | 6.132 |
| PdMnSn | 6.200 |

*Table 2 Lattice Parameters of Full Heusler Alloys*

| $L2_1$ Alloy | Lattice Parameter (Å) |
|:---:|:---:|
| $Au_2MnSn$ | 6.591 |
| $Cu_2NiGe$ | 5.851 |
| $Pd_2NiGe$ | 6.200 |
| $Pt_2CoSn$ | 6.342 |

Among the Half-Heusler alloys mentioned in the table 1, AuCrSn had the largest lattice parameter due to the presence of larger atoms Au and Sn in the lattice sites X and Z of XYZ composition respectively. The lattice parameter of PdMnSn in the relaxed geometry was found to be 6.200 Å, which is in good agreement with Nihat et al. [3]'s assessment of lattice parameters and electronic structure of PdMnSn. In their work the authors used DFT with generalized gradient approximation (GGA) to attain the relaxed geometry and observed that the DFT results had tendency to underestimate the lattice parameters of the alloys [3]. As the alloys considered in this work except PdMnSn are novel materials, there is no published literature on the ab initio studies or experimental studies on the alloys AuCrSn, AuMnGe, $Au_2MnSn$, $Cu_2NiGe$, $Pd_2NiGe$ and $Pt_2CoSn$.

## 3.2 Electronic Structure

In this section, the electronic structures of both the families of Heusler alloys are discussed. Due to the differences in the lattice structure of $C1_b$ and $L2_1$ alloys, the origin of bandgap in the minority spin of band structures for both families are different. These mechanisms, which cause the bandgap are discussed. The presence or absence of bandgap is explained based on the bandgap origins presented by *Galanakis et al.* [1]. In addition to the band structure, localized and total magnetic moment of the lattice is calculated for the seven alloys studied in this work.

### 3.2.1 Origin of Bandgap in Half-Heusler Alloys

As is the case for many Half-Heusler alloys such as NiMnSb and CoTiSb, there exists a bandgap in the band structure of minority configuration [1]. This characteristic of the materials is also called half-metallicity. This bandgap arises from the hybridization between the d orbitals of the atoms located in X and Y lattice sites of the $C1_b$ phase (XYZ) unit cell. The density of states (DOS) of widely studied alloy NiMnSb shows a bandgap in the minority configuration in Figure 3.

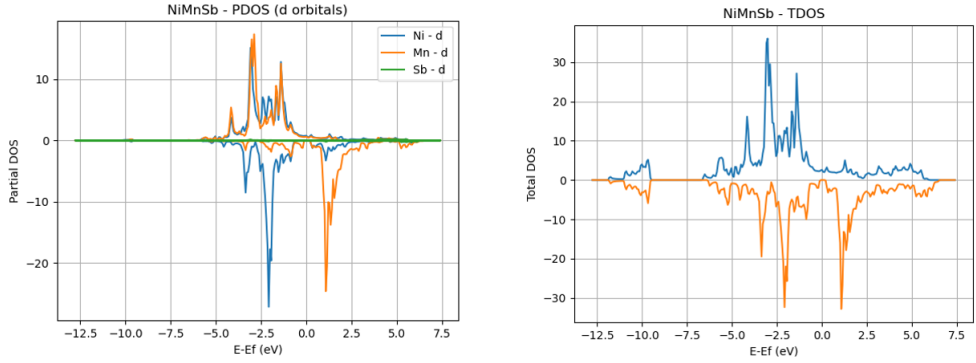

*Figure 3 NiMnSb - Partial and Total Density of States. The PDOS of d-orbital projection have been shown in the left plot.*

It can be observed in the PDOS of NiMnSb in Figure 3 that the d bands of Ni and Mn atoms have hybridized. Due to the hybridization, between the d-bands of the Ni and Mn, the majority Mn-d bands shifts to the lower energies and form hybridized orbitals with Ni [1]. The occupied majority hybrid d states sit below $E_F$ and unoccupied high energy minority hybrid d states are shifted above the $E_F$. Figure 4 showcases the hybridization between the d-d orbitals of Ni and Mn atoms [1]. This phenomenon causes the bandgap that sits across $E_F$ in NiMnSb and similar half-metallic alloys crystallizing in C1$_b$ structure, according to the work of *Galanakis et al* [1].The valence band is dominated by the Ni (X) admixture and the conduction band is dominated by the Mn (Y) admixture of antibonding hybrids [1]. Hence the strong d-d hybridization of X and Y species in C1$_b$ structure causes the bandgap [1]. The Z species (sp element) does not contribute to the cause of bandgap; however, it is essential for the stability of the compound due to its electronegativity [1].

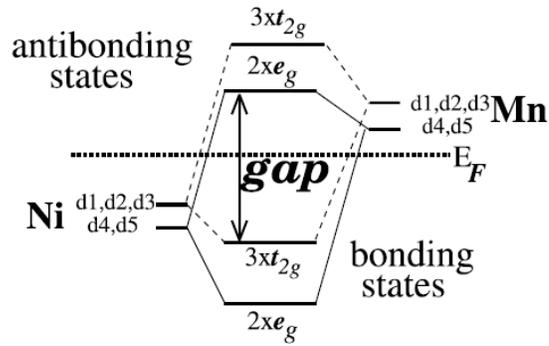

*Figure 4 d-d Hybridization of the Ni and Mn in NiMnSb alloy [1]*

As the X & Y species are responsible for the existence of bandgap and half-metallic character of the alloy, their atomic positions in the lattice, which are arranged in the zinc-blende structure are crucial as it allows the specific bonding and hybridization of the d bands [2]. Reader is referred to [2] for further explanation of d-d hybridization.

### 3.2.2 Electronic Structure of Half-Heusler Alloys

**AuCrSn**

AuCrSn alloy crystallizes in the $C1_b$ phase with Au, Cr and Sn in the X, Y and Z positions of $C1_b$ structure shown in Figure 1. The valence electron configurations of Au, Cr and Sn are $5d^{10}6s^1$, $3d^54s^1$ and $5s^25p^2$ respectively. The partial DOS, total DOS as well as spin polarized band structure of AuCrSn alloy are shown in Figures 5 , 6 and 7 respectively. The s bands originating from the Sn atoms sit down low in the region of -10 to -7 eV and the p bands are spread out across the spectrum. The d bands of Sn atoms are fully occupied, and they do not participate in any bonding. The 5d bands of Au are also fully occupied, so they sit at a much lower energy level between -6 to -4 eV, away from the $E_F$ and Cr-d states above -2 eV and spans across $E_F$. There is no overlap between the Au and Cr-d bands. Due to this reason, there is no hybridization between Au-d and Cr-d bands. Hence, there exists no bandgap at $E_F$ in minority band structure.

In AuCrSn, Au-d bands being completely occupied, localized magnetic moment of Au sites obtained from ab initio calculations is -0.011 $\mu_b$. The localized magnetic moment of Cr and Sn was calculated to be 3.634 $\mu_b$ and -0.182 $\mu_b$. The calculated total magnetic moment of the unit cell is 3.441 $\mu_b$. The magnetic moments calculated in this work are in the unit of Bohr magnetons ($\mu_b$) [21] per atom. One Bohr magneton can be defined by the term $\frac{e\hbar}{2m_e}$, where e is the electron charge, $\hbar$ is the reduced Planck constant and $m_e$ is the electron mass [21]. The negative magnetic moments of Au and Sn imply that the Au and Sn atoms have an antiferromagnetic correlation with Cr atoms. In the Heusler alloy family, low magnetic moment of the material helps preserve the half-metallicity of the alloy [1]. Hence, with the rise in total magnetic moment of the unit cell, half-metallicity is not retained. AuCrSn is not half-metallic as there exists no bandgap in the band structure at $E_F$.

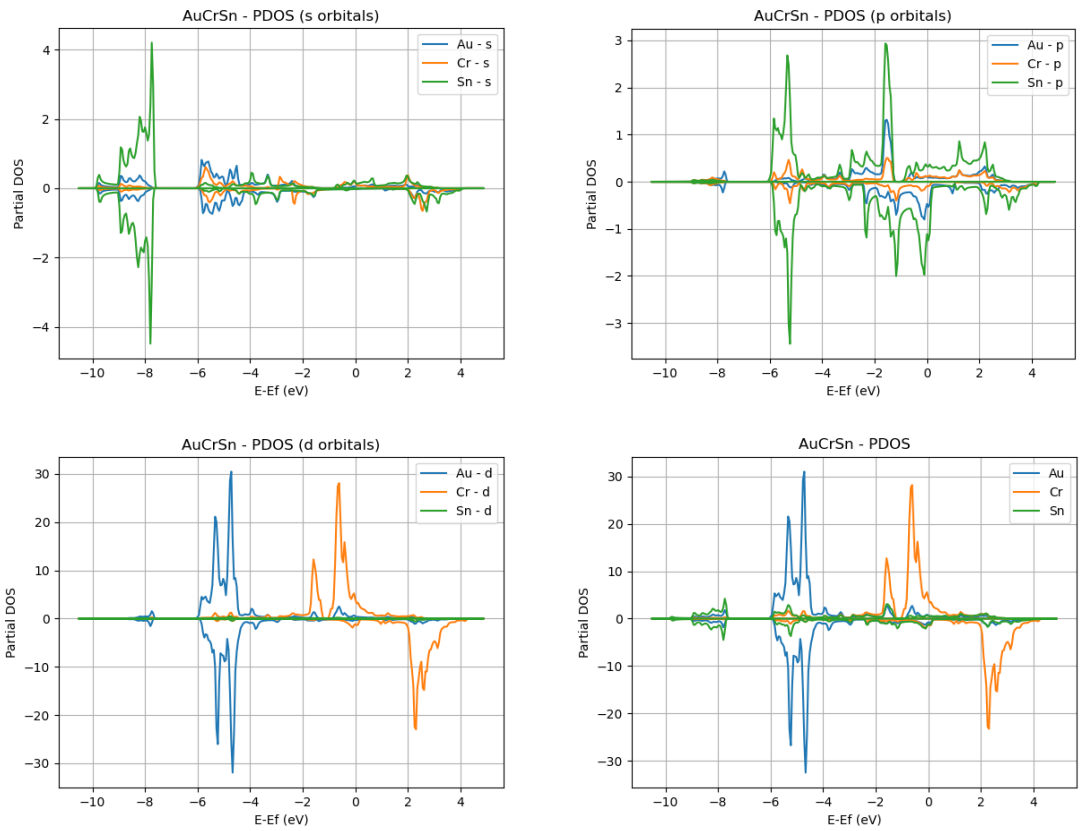

*Figure 5 AuCrSn - Partial Density of States*

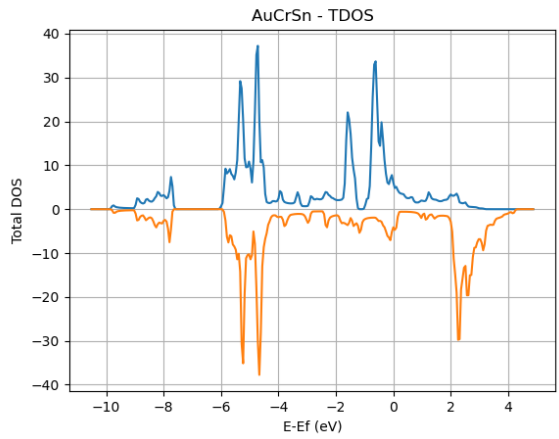

*Figure 6 AuCrSn - Total Density of States*

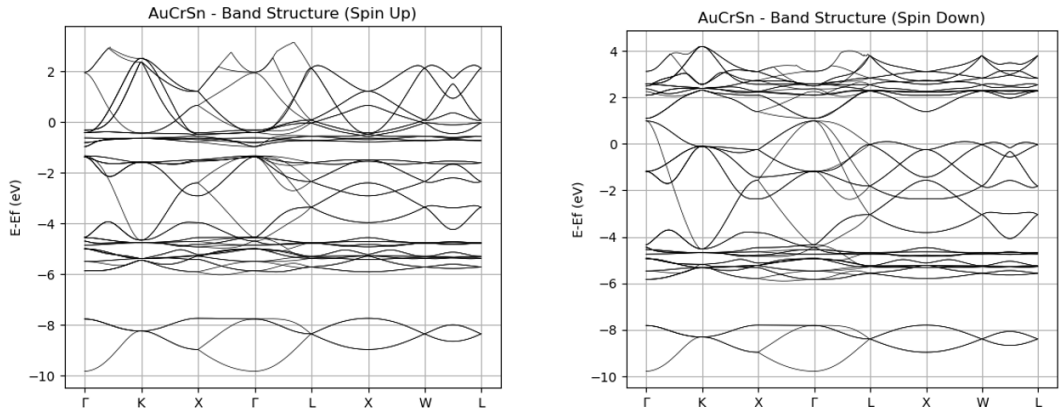

*Figure 7 AuCrSn - Band Structure*

**AuMnGe**

As described in the electronic structure of AuCrSn, the s bands of the sp element of AuMnGe are at the lowest of the energy levels, in the range of -11 to -8 eV, while the p bands span across the energy range of -7 to +6 eV. The valence electron configuration of Au, Mn and Ge is $5d^{10}6s^1$, $3d^54s^2$ and $4s^24p^2$ respectively.

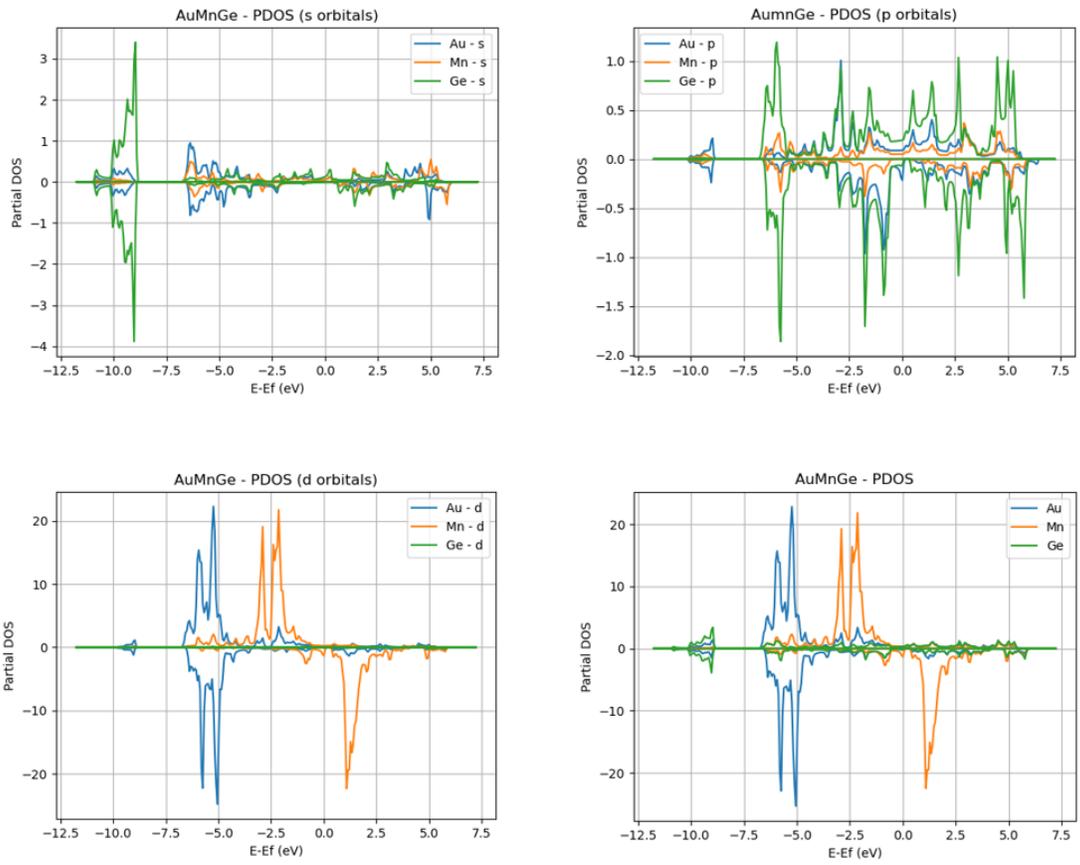

*Figure 8 AuMnGe - Partial Density of States*

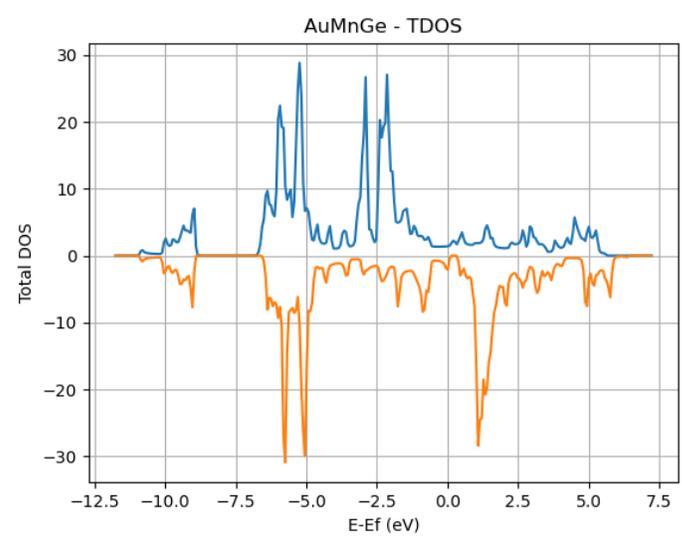

*Figure 9 AuMnGe - Total Density of States*

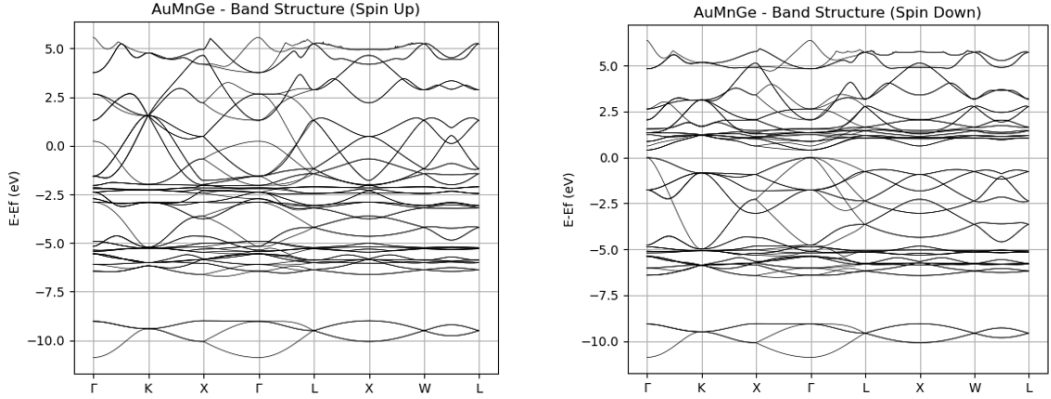

*Figure 10 AuMnGe - Band Structure*

As the Figures 8,9, and 10 indicate, and observed in the case of AuCrSn, the d bands of Au atoms sit in the low energy region of -7.5 to -4 eV, which is due to all 5 Au-d states being fully occupied. The Mn-d states are located much closer to the $E_F$, with majority states being dominant in the valence band and minority states dominating in the conduction band. In AuMnGe, a small overlap of d bands of Au and Mn can be observed. This overlap leads to the existence of a small minority bandgap, beginning at $E_F$, expanding into the conduction band. Hence AuMnGe showcases half-metallic behavior.

The calculated localized magnetic moment for Au, Mn and Ge sites is 0.045 $\mu_b$, 3.910 $\mu_b$ and -0.083 $\mu_b$. The total magnetic moment of the unit cell is 3.872 $\mu_b$. Unlike AuCrSn, Au atom in X site has ferromagnetic correlation with Mn atom in Y site. The total magnetic moment of the unit cell is 3.872 $\mu_b$.

**PdMnSn**

In PdMnSn alloy crystalizing in C1$_b$ structure, the valence electron configuration of its constituent elements Pd, Mn and Sn is $4d^{10}$, $3d^5 4s^2$ and $5s^2 5p^2$ respectively. The electron structure of PdMnSn differs significantly from that of AuCrSn and AuMnGe. PdMnSn has deep-lying s bands from sp element occupying Z sites.

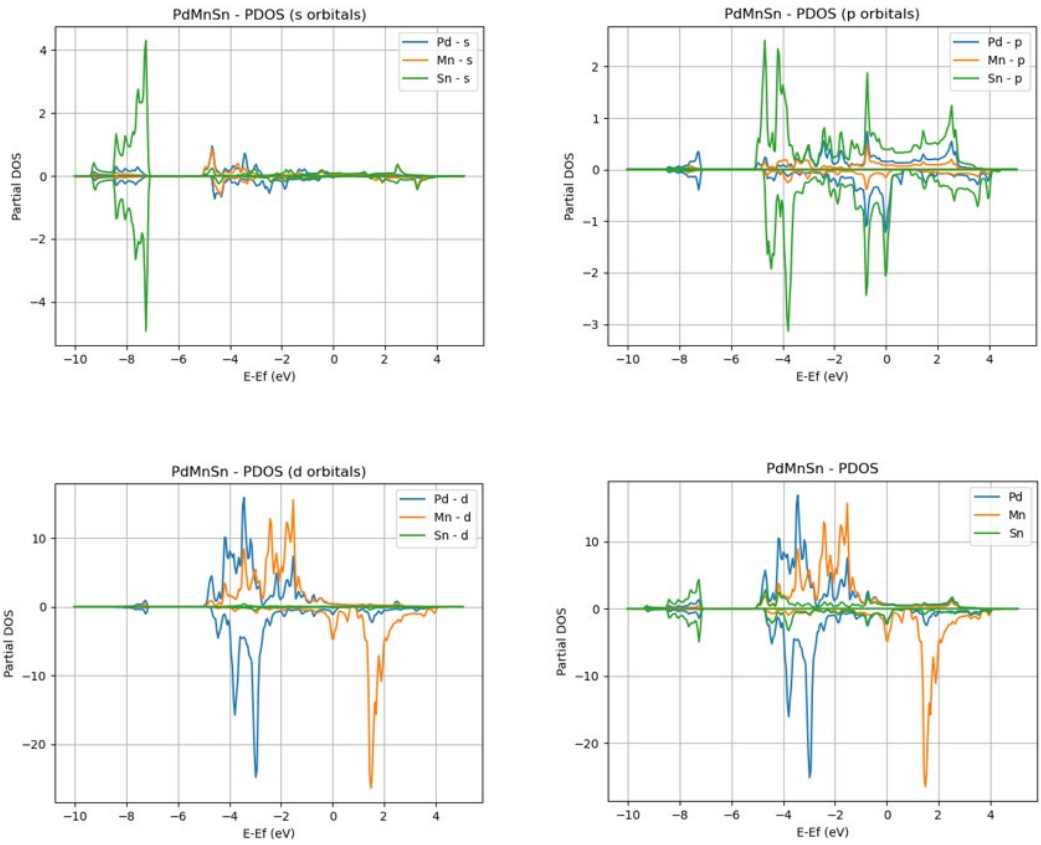

*Figure 11 PdMnSn - Partial Density of States*

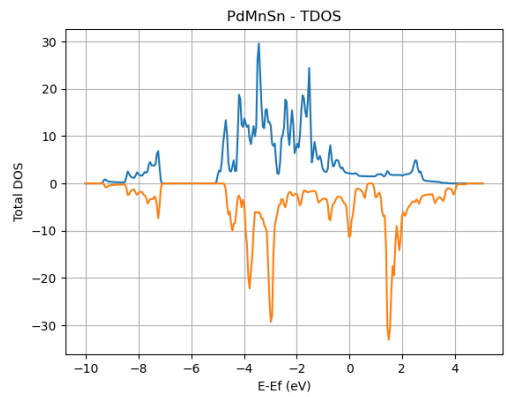

*Figure 12 PdMnSn - Total Density of States*

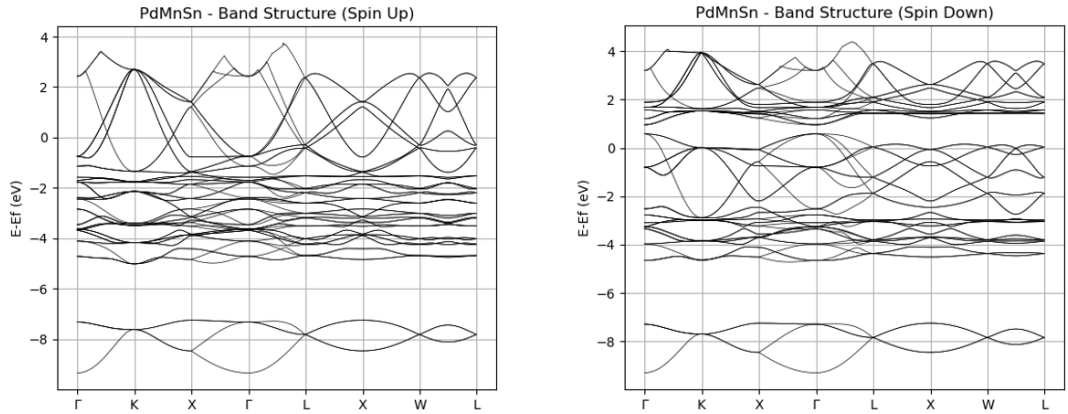

*Figure 13 PdMnSn - Band Structure*

These s bands are located in the proximity of -8 eV and the p bands span across the spectra. Figures 11, 12 and 13 show the partial DOS, total DOS and band structure of PdMnSn respectively. The d bonds of Pd atom's valence shell fully occupied and these d bands are located in the energy range of -6 eV to $E_F$. The majority d-bands (spin up) of Mn are fully occupied, however, all minority d-bands are empty. The majority Mn-d bands are located in the valence bands and minority Mn-d bands are spread into the conduction bands. There are little to no signs of d-d hybridization between the Pd (X sites) and Mn (Y sites) elements, due to d orbital of Pd being fully occupied. Hence no bandgap was found in PdMnSn. PdMnSn indicated no characteristics of half-metallicity. These results are consistent with the work of Nihat et al. [3]

The localized magnetic moments obtained from ab initio calculations of Pd, Mn and Sn occupying X, Y and Z sites are 0.056 $\mu_b$, 3.908 $\mu_b$ and -0.151 $\mu_b$ respectively. Total magnetic moment of the PdMnSn unit cell is 3.813 $\mu_b$. Pd and Mn atoms are ferromagnetically correlating as both the atoms have positive localized magnetic moment.

### 3.2.3 Origin of Bandgap in Full-Heusler Alloys

Many Full-Heusler alloys have been extensively studied and found to have the minority bandgap in their electron structure, characterizing half-metallic behavior [1]. Alloys like $Co_2MnSb$ and $Co_2MnGe$ have been found to have bandgap in minority spin configuration of band structure [1]. Full-Heusler alloys of $X_2YZ$ composition crystallize in $L2_1$ structure. The $L2_1$ structure is shown in Figure 1. The $L2_1$ and $C1_b$ structures are very similar to each other and $C1_b$ structure can be obtained by eliminating 4 atoms from the $X_2$ sites. Due to their similar lattice structure, it would be fair to assume that the origin of the bandgap in Full-Heusler alloys would be similar to that of $C1_b$ alloys.

In $L2_1$ structure, two X atoms are located at the second neighbor distances in the lattice and Y atoms are in the body centered sites. This arrangement allows the Y atoms to have 8 X atoms in the direct neighboring positions. However, the primary hybridization that takes place in $L2_1$ phase alloys is between the d orbitals of two X atoms. This hybridization creates two-fold degenerates $2 \times e_g$ & $2 \times e_u$ and three-fold degenerates $3 \times t_{2g}$ & $3 \times t_{1u}$ hybridized orbitals [1]. The $e_g$ and $t_{2g}$ are the bonding orbitals, which have lower energy and $e_u$ and $t_{1u}$ are antibonding orbitals with higher energy [1]. Figure 14 illustrates the hybridized bands in $Co_2MnSb$ [1].

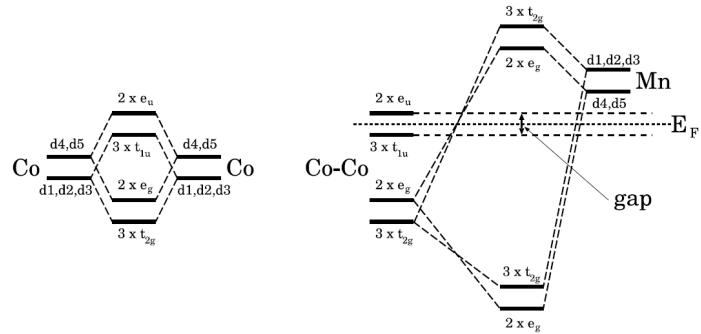

*Figure 14 Hybridization in the Full-Heusler Alloy $Co_2MnSb$ [1]*

These hybridized orbitals then engage in the second step of hybridization with d states of Y atoms. The $e_g$ orbitals hybridize with the $d_{z^2}$ and $d_{x^2-y^2}$ of Y atom and the hybrid $t_{2g}$ further hybridizes with $d_{xy}$, $d_{yz}$, and $d_{zx}$ of Y atom. The splitting of bonding and antibonding hybrids from the primary d-d hybridization between the X atoms causes the bandgap at $E_F$ [1].

### 3.2.4 Electronic Structure of Full-Heusler Alloys

**Au₂MnSn**

The density of states for $Au_2MnSn$ spread in the energy range of -12.5 to +7.5 eV. Au, Mn, and Sn occupying X, Y and Z sites have valence electron configure of $5d^{10}6s^1$, $3d^54s^2$, and $5s^25p^2$ respectively. Similar to the Half-Heusler alloys, the band structure of $Au_2MnSn$ has deep lying s bands from the sp element Sn. In addition to the 5s bands from Sn, 6s bands of Au are also there, which spreads into the conduction band. As there are twice as many X atoms in the unit cell compared to Y and Z, the Au-s and Au-d states have higher density. All the Au-d bands are fully occupied, so Au-d bands are at lower energy levels. Because of Au-d bands being fully occupied, d-d hybridization between Au-d bands is not possible. This further eliminates the possibility of hybridization with Mn-d orbitals. Due to this reason, there is no minority bandgap at $E_F$ in $Au_2MnSn$. Due to the absence of minority bandgap, it can be concluded that $Au_2MnSn$ is not half-metallic.

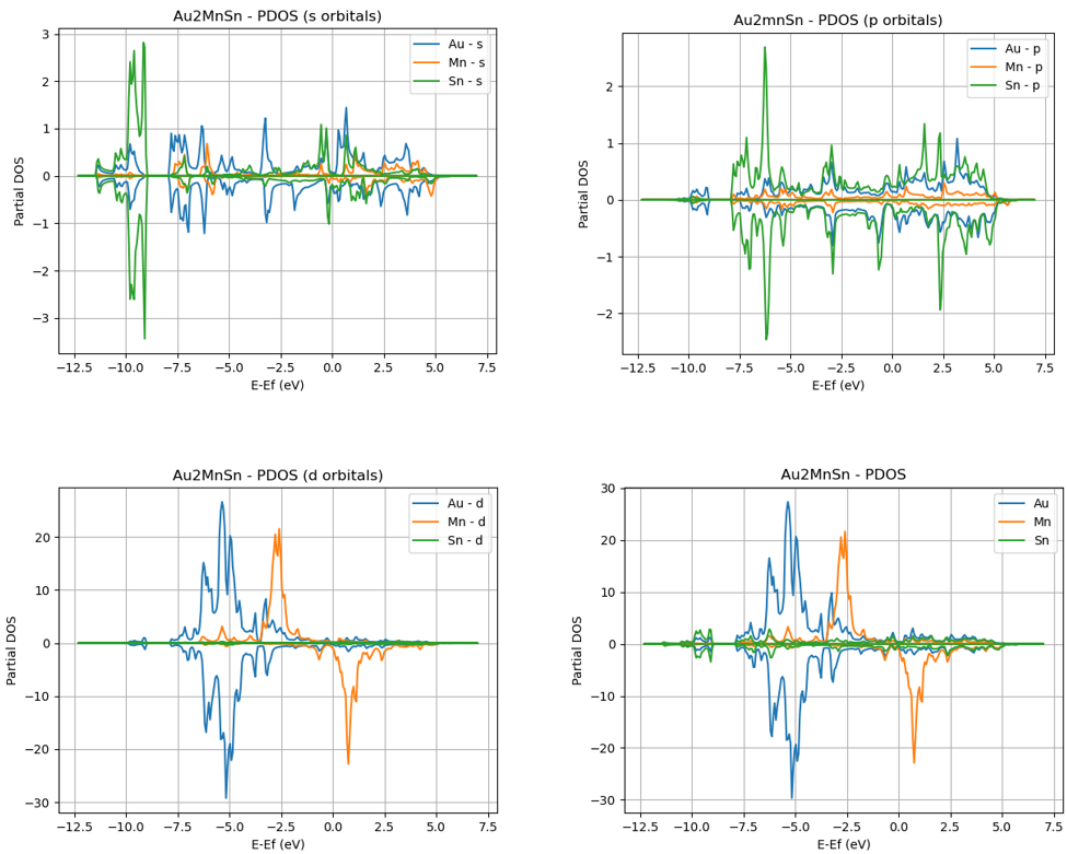

*Figure 15 Au₂MnSn - Partial Density of States*

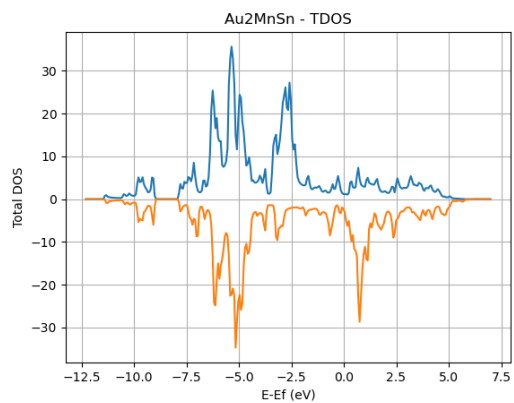

*Figure 16 Au₂MnSn - Total Density of States*

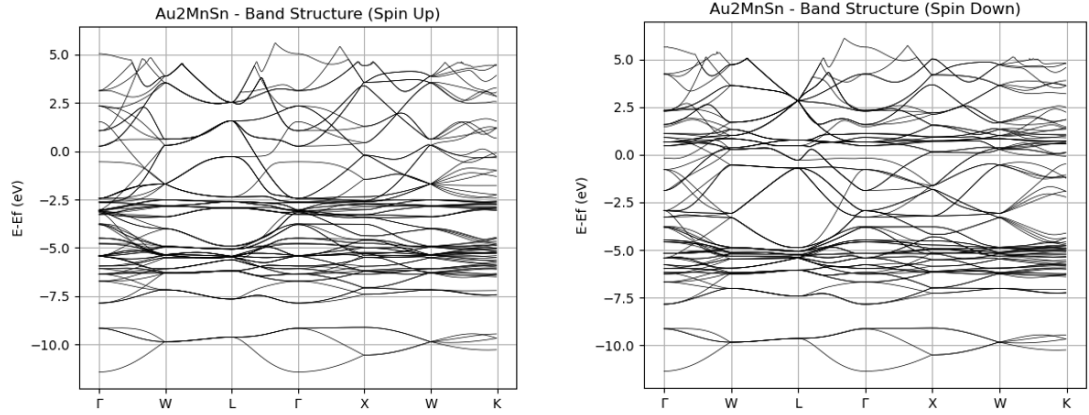

*Figure 17 Au₂MnSn - Band Structure*

The magnetic moment in Full-Heusler alloys is concentrated on the Y atoms like Half-Heusler alloys. In $Au_2MnSn$ unit cell, localized magnetic moment of Au, Mn and Sn atoms are -0.006 $\mu_b$, 3.825 $\mu_b$ and 0.0046 $\mu_b$ respectively. The Au and Mn atoms have antiferromagnetic correlation.

**$Cu_2NiGe$ and $Pd_2NiGe$**

The electron structure of $Cu_2NiGe$ and $Pd_2NiGe$ has been found to be the most unique among the seven alloys studied in this work. These two alloys share the same elements in Y and Z body centered positions in the lattice structure. All the half-metallic Heusler alloys tend to have 100% spin polarization of valence shell. The semiconducting materials of the Heusler family do not have spin polarization of valence shell. Figures 18-21 suggest that the electron structures of $Cu_2NiGe$ and $Pd_2NiGe$ are not spin polarized.

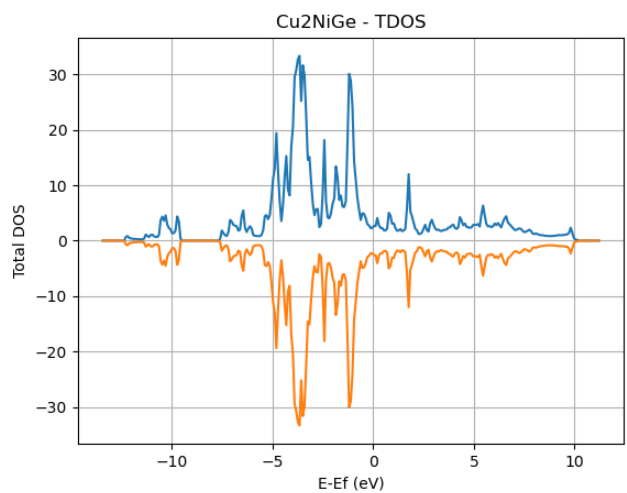

*Figure 18 Cu₂NiGe - Total Density of States*

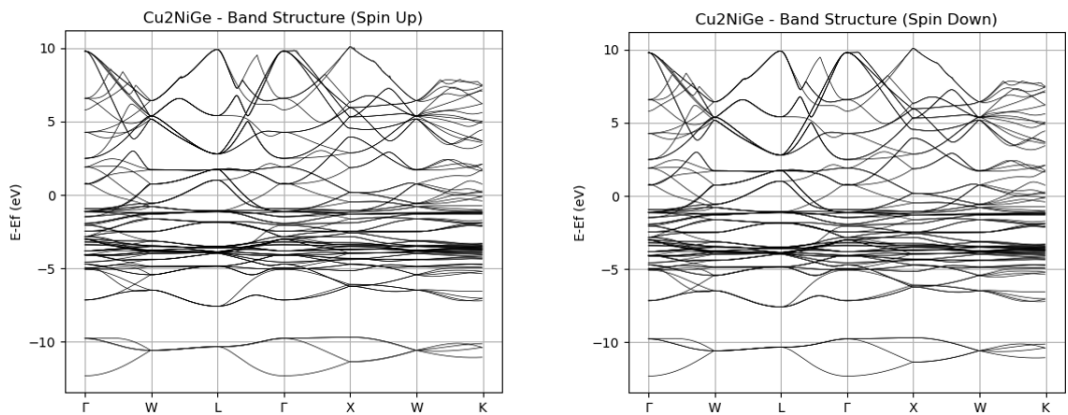

*Figure 19 Cu₂NiGe - Band Structure*

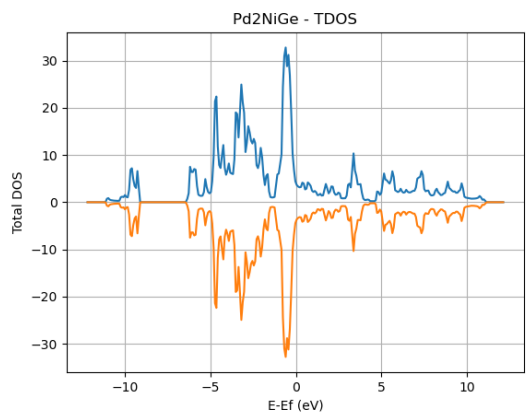

*Figure 20 Pd₂NiGe - Total Density of States*

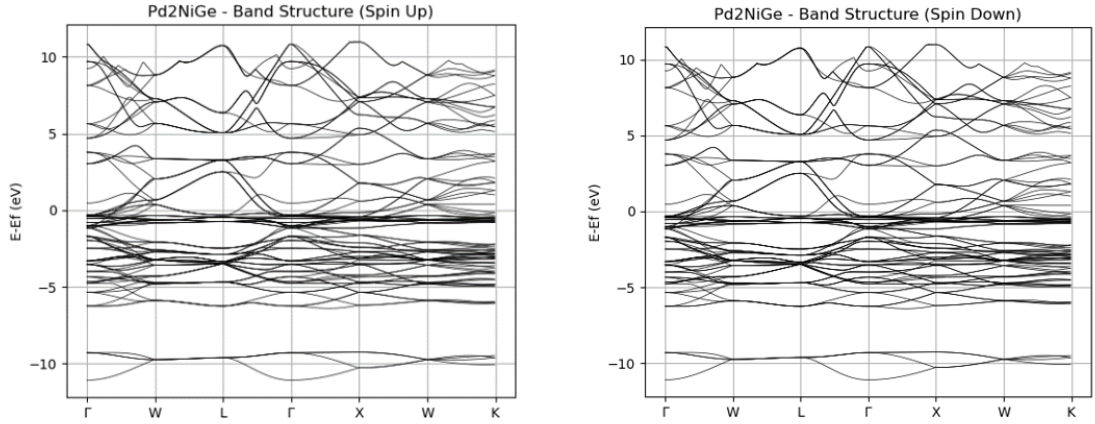

Due to absence of spin polarization, the discussion of bandgap for these two materials is futile and it is evident that these alloys do not exhibit half-metallicity. As the materials are not spin polarized and with Ni and Ge atoms in Y and Z positions, both materials showcase metallic characteristics. The similarities in the DOS of both materials can be drawn from figures 18 and 20. The conduction bands of both materials have similar densities due to common elements Ni and Ge in the Y atomic sites. Total and localized magnetic moment calculated at all sites in $Cu_2NiGe$ and $Pd_2NiGe$ is 0 $\mu_b$. Because of the absence of spin polarization, which is the primary indicator of half-metallicity, it can be safely concluded that these two materials are not semi-metallic. Hence, the PDOS plots have not been incorporated for these two materials.

**$Pt_2CoSn$**

As shown in figures 22, 23 and 24, the valence electrons of $Pt_2CoSn$ were found to be spin polarized like other alloys from Heusler family. The valence electron configuration of the constituent atoms Pt, Co and Sn are $5d^96s^1$, $3d^74s^2$ and $5s^25p^2$ respectively.

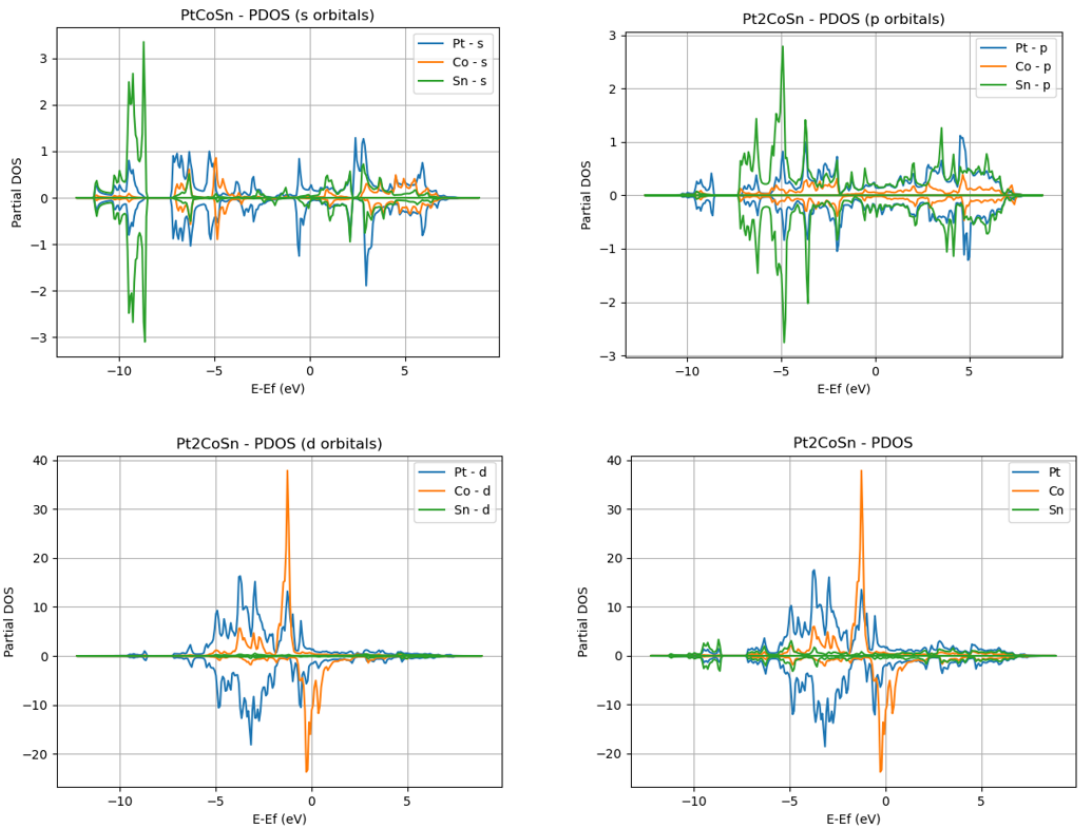

*Figure 22 Pt₂CoSn - Partial Density of States*

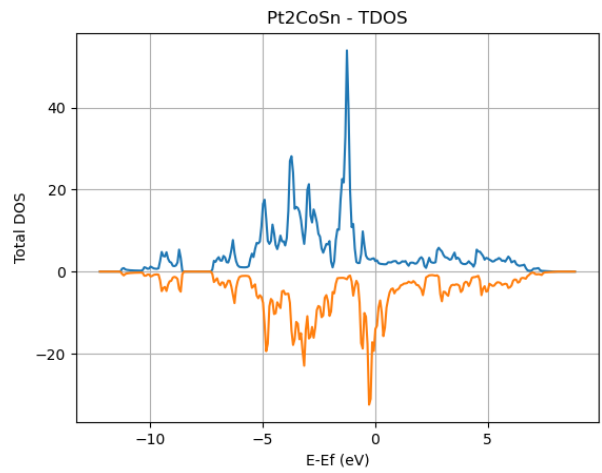

*Figure 23 Pt₂CoSn - Total Density of States*

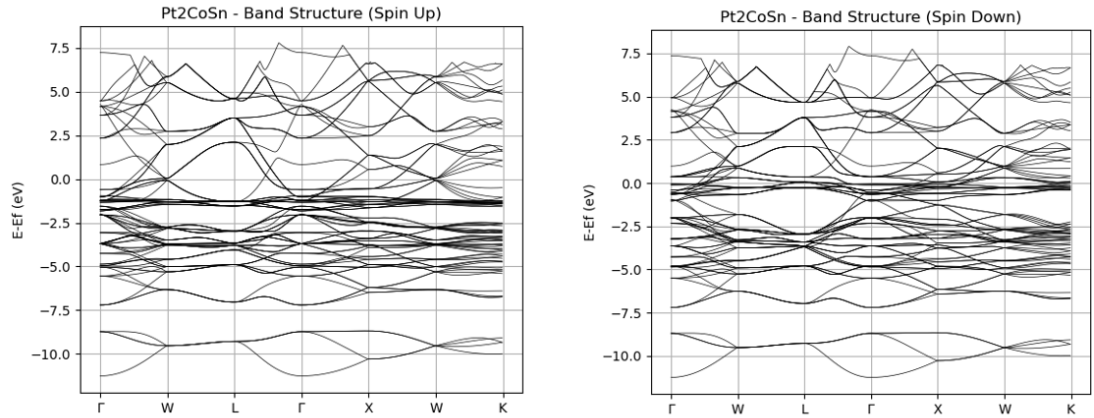

*Figure 24 Pt₂CoSn - Band Structure*

As observed in the 6 alloys previously mentioned in this work, sp element provides the deep-lying s bands at lower energies. Pt provides the d-bands. These Pt-d bands are fully occupied in majority spin and one minority band is unoccupied. The majority Co-d admixture dominates the valence bands, while minority d bands expand into the conduction band. Due to only one unoccupied Pt-d band, no d-d hybridization is possible. It eliminates the further hybridization with Co-d bands and the bandgap in minority spin ceases to exist at $E_F$. Pt₂CoSn is also not a half-metallic alloy.

The magnetic moment of unit cell was calculated to be 1.825 $\mu_b$, with localized magnetic moments of Pt, Co and Sn being 0.064 $\mu_b$, 1.765 $\mu_b$ and -0.004 $\mu_b$ respectively. Positive magnetic moment of Pt and Co indicate the ferromagnetic correlation between X-Y sites.

### 3.3 Phonon Spectra

### 3.3.1 Phonon Spectra of Half-Heusler Alloys

**AuCrSn**

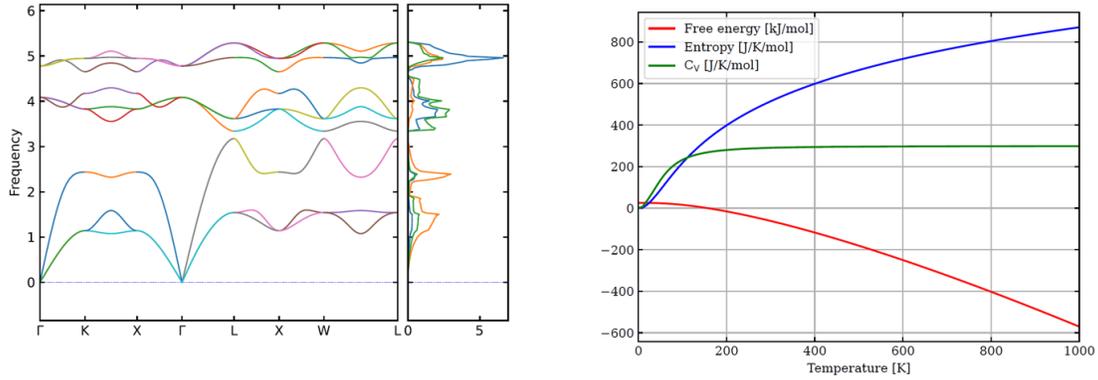

*Figure 25 AuCrSn - Phonon modes (THz) and Thermal Properties*

The phonon spectra for all the seven alloys have been calculated with finite difference method [15], as mentioned in the methodology section. The Half-Heusler alloys of $C1_b$ structures have been observed to have 9 phonon modes across the high symmetry path defined for $C1_b$ structures. The phonon band plot in Figure 25 shows the phonon modes (THz) against the high symmetry path for AuCrSn.

AuCrSn has 9 phonon modes. This phonon spectra consists of 3 acoustic modes and 6 optical modes that are tightly dispersed and can be observed in the frequency range of 0-3 THz. The majority of high frequency optical phonon modes come from the Cr atoms in Y positions of the lattice followed by the vibrational modes due to the heavier atoms Au and Sn. The acoustic modes in AuCrSn are dominated by the vibrations of Au atoms. In current study of the phonon modes, no imaginary phonon modes were observed for AuCrSn. It indicates that the AuCrSn crystal structure is thermodynamically stable in $C1_b$ structure and there are no vibrations that may introduce the instability to the $C1_b$ phase of AuCrSn. Figure 25 also showcases the variation of thermal properties Gibbs Free energy (G), entropy (S) and specific heat at constant volume (C$_v$) with temperature in the range of 0-1000 K. From absolute zero temperature, C$_v$ rises rapidly up to 300 K and shows asymptotic behavior.

**AuMnGe**

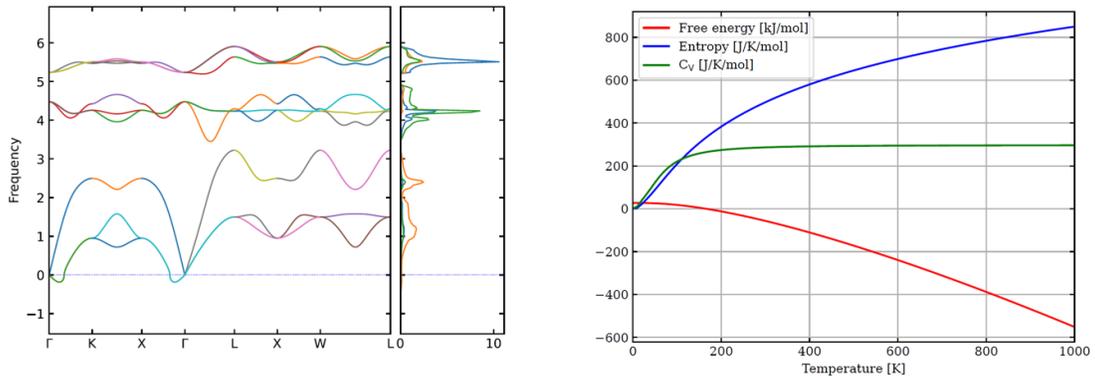

*Figure 26 AuMnGe - Phonon Modes (THz) and Thermal Properties*

The phonon study of the AuMnGe gives the phonon modes as shown in figure 26. It can be observed in the phonon spectra of AuMnGe that there are 9 phonon modes, 3 of which are optical phonon modes, with frequencies lower than 3.5 THz and originates from the gamma point of reciprocal lattice. The optical phonon modes can be observed in the two groups of 3 modes each originating between the frequencies of 4-6 THz. As is the case for AuCrSn, the high frequency optical phonon modes are dominated by the vibrations of relatively smaller Mn atoms in Y sites, with relatively low contribution from the vibrations of heavier atoms in X and Z positions. The low frequency optical phonon modes come from the vibration of the sp element Ge. As observed in AuCrSn, the acoustic modes are dominated by the Au vibrations.

The vibrations of negative (imaginary) frequencies suggest that their amplitudes increase instead of damping out in the lattice. These vibrations are unsustainable as it does not allow the lattice structure to be stable. It can be observed that the acoustic modes have a small imaginary component to their vibrational frequencies, arising from the unsustainable vibrations from heavier Au atoms. It introduces the thermodynamic instability to the C1$_b$ structure of AuMnGe. Figure 26 shows the calculated thermal properties (G, S, and C$_v$) of AuMnGe as a function of temperature in the temperature range of 0-1000 K. However, due to the presence of negative phonon frequencies, the thermal properties G, S, and C$_v$ are calculated based on the real component of phonon modes only and does not represent the actual behavior of the AuMnGe.

**PdMnSn**

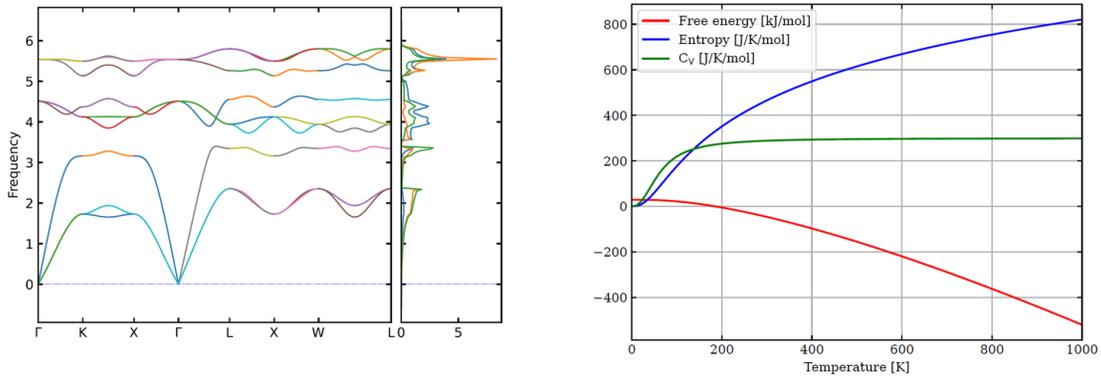

*Figure 27 PdMnSn - Phonon Modes (THz) and Thermal Properties*

Figure 27 shows the phonon modes (THz). The phonon spectra of PdMnSn ranges from 0-6 THz. It consists of 9 phonon modes, 3 of them being acoustic and the rest of the 6 modes being optical, it confirms to the C1$_b$ structure. The high wavelength and low frequency acoustic phonon modes have frequencies in the range of 0-3.5 THz and the optical phonon modes exist in the two groups. Each group has 3 phonon modes. Unlike

AuCrSn and AuMnGe, the high frequency optical modes have major contribution from the Pd atoms in X sites. These results can be justified by the smaller size of the Pd atoms compared to Au. The contribution of Mn and Sn vibrations to these modes is fairly equal. The low energy optical modes in the range of 3.5-5 THz have major contribution from the vibrations of Sn and so is the case for acoustic modes below 3.5 THz.

From the observations made in this work, it is hypothesized that due to the similar size of all three constituent elements in PdMnSn, no imaginary phonon modes were observed. Hence it can be concluded that the PdMnSn is thermodynamically stable in C1$_b$ phase. These results are consistent with the work of Nihat et al. [3]. Figure 27 shows the thermal properties G, S and Cv of PdMnSn as a function of temperature, calculated based on the phonon modes of the system. As the calculated phonon spectra of PdMnSn does not have any negative modes, the calculated thermal properties (G,S and C$_v$) are truly representing the PdMnSn.

### 3.3.2   Phonon Spectra of Half-Heusler Alloys

**Au$_2$MnSn**

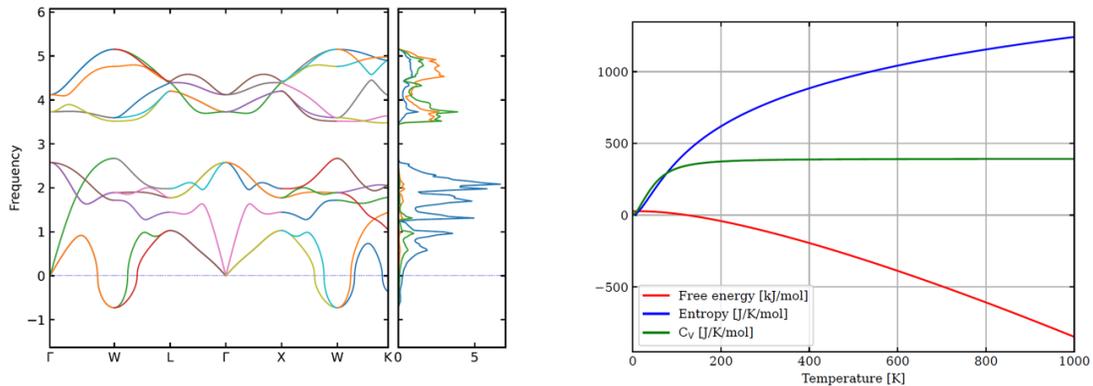

*Figure 28 Au$_2$MnSn - Phonon Modes (THz) and Thermal Properties*

The phonon modes of $Au_2MnSn$ are shown in figure 28. Full-Heusler alloys with $L2_1$ crystal structure have been studied to have 12 phonon modes, 3 modes arising from each atom of $X_2YZ$ composition. The phonon spectra consist of 9 optical phonon modes and 3 acoustic modes. The optical modes exist in 3 groups of 3 modes each. The acoustic modes of $Au_2MnSn$ have frequencies lower than 3 THz. These acoustic phonon modes are dominated by the vibrations of heavy Au atoms in $X_1$ and $X_2$ sites. It can be observed that the contribution of Mn and Sn vibrations towards acoustic modes is very low compared to Au and so is the case for low energy optical phonon modes lying below 3 THz. The high frequency optical modes originating above 4 THz have major contribution from the Mn atoms at Y sites and three optical modes originating between 3-4 THz are dominated by the vibrations of sp element Sn.

It is evident from the phonon plot that the acoustic modes have imaginary vibrational frequencies arising from the heavier atomic sites of Au in the lattice. These unsustainable Au vibrations introduce thermodynamic instability to the $L2_1$ phase of $Au_2MnSn$. Figure 28 shows the thermal properties (G, S and $C_v$) of $Au_2MnSn$ as function of temperature in the temperature range of 0-1000 K. As the phonon spectra of $Au_2MnSn$ consists of imaginary phonon modes, the calculated thermal properties are not truly representative of $Au_2MnSn$ alloy.

**Cu₂NiGe**

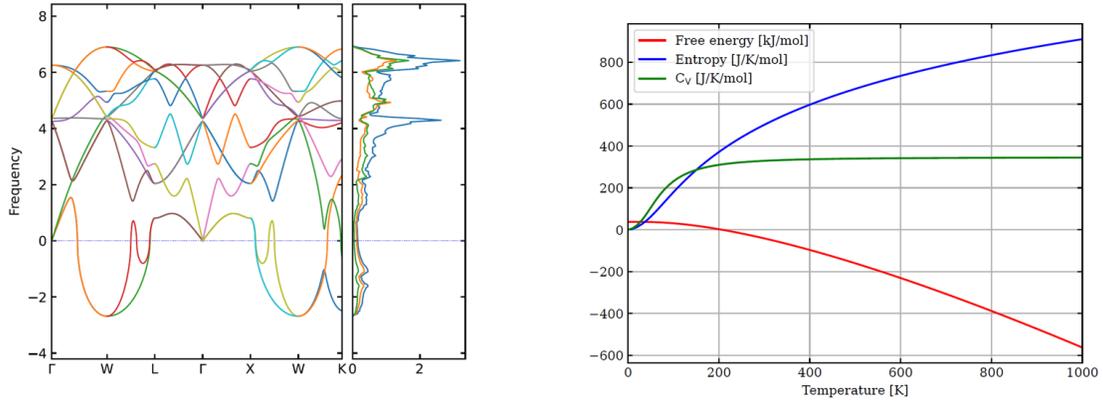

*Figure 29 Cu₂MniGe - Phonon Modes (THz) and Thermal Properties*

The Phonon modes of Cu₂NiGe are shown in figure 29 along with the PDOS. Cu₂NiGe being a Full-Heusler alloy with L2₁, it has 12 phonon modes, consisting of 3 acoustic and 9 optical modes of lattice vibration. In the case of three Half-Heusler alloys and Au₂MnSn, the optical modes were observed to be isolated from acoustic modes on phonon spectra. This is not the case for Cu₂NiGe. The modes are widely dispersed and not isolated. All 9 optical modes originate between 5-6.5 THz, and they are dominated by Cu vibrations. The high frequency modes have not been dominated by the X atoms in the alloys previously studied in this work due to their larger size. As the X sites are occupied by much lighter Cu atoms (compared to Au and Pd), the high frequency vibrations of X atoms are justified. The acoustic modes originating from gamma point consist of imaginary vibrational frequencies and it is evident that the highest contribution to acoustic modes is from the sp element Ge, which is the heaviest element in the composition of this alloy. The non-isolation of phonon modes can be attributed to the similar sizes of the atoms, causing them to vibrate at similar frequencies.

Due to the unsustainable vibrations of heavy Ge atoms in body centered Z sites, the thermodynamic instabilities are introduced in the L2$_1$ phase of Cu$_2$NiGe. Figure 29 shows the thermal properties (G, S and C$_v$) of Cu$_2$NiGe as a function of temperature in the temperature range of 0-1000 K. As the phonon spectra of Cu$_2$NiGe consists of imaginary phonon modes, the calculated thermal properties are not truly representative of Cu$_2$NiGe alloy.

**Pd$_2$NiGe**

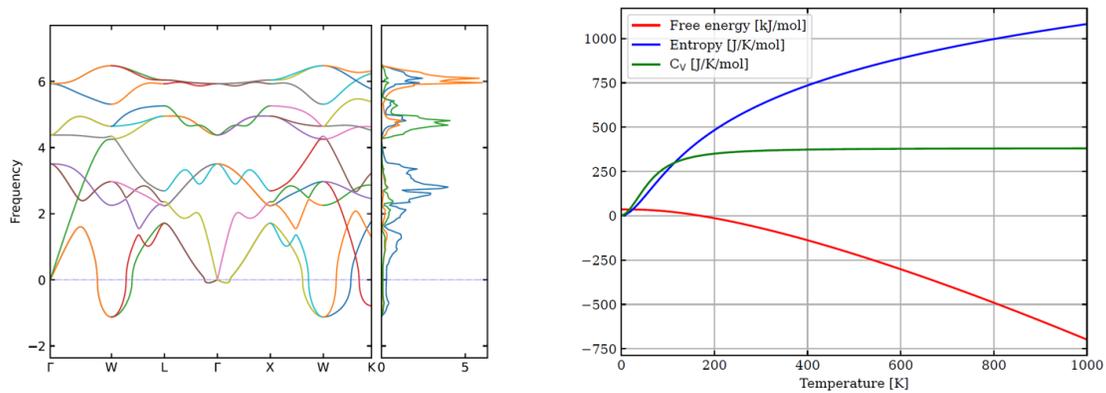

*Figure 30 Pd$_2$NiGe - Phonon Modes (THz) and Thermal Properties*

Phonon modes of Pd$_2$NiGe along with PDOS are shown in figure 30. Pd$_2$NiGe being a Full-Heusler alloy with L2$_1$ structure, has 12 phonon modes. These modes are dispersed across the frequency range of -1 to 7 THz. The optical modes are separated into three groups of three phonon modes each. Pd$_2$NiGe has disparity in the size of atoms occupying X, Y and Z sites lattice, allowing the isolation of phonon modes. The high frequency optical modes originating at 6 THz are dominated by the lighter Ni atoms in Y sites, and least contribution is from the heavier Ge atoms. The lower frequency phonon modes are dominated by the vibrations of heavier atoms and vice versa. The low frequency optical modes and acoustic modes ae dominated by the Pd vibrations.

The Pd vibrations contribute the most towards the imaginary frequencies of acoustic phonon modes. These unsustainable Pd vibrations introduce thermodynamic instability to the L2$_1$ phase of Pd$_2$NiGe. Figure 30 shows the thermal properties (G, S and C$_v$) of Pd$_2$NiGe as a function of temperature in the temperature range of 0-1000 K. As the phonon spectra of Pd$_2$NiGe consists of imaginary phonon modes, the calculated thermal properties are not truly representative of Pd$_2$NiGe alloy.

**Pt$_2$CoSn**

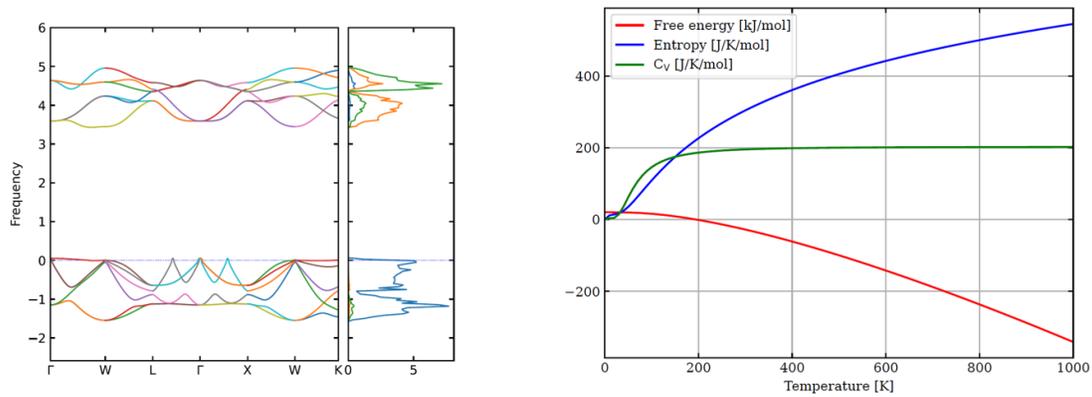

*Figure 31 Pt$_2$CoSn - Phonon Modes (THz) and Thermal Properties*

Phonon modes of Pt$_2$CoSn along with PDOS are shown in Figure 31. Similar to the phonon spectra of other Full-Heusler alloys, Pt$_2$CoSn has 12 phonon modes, 3 of them being acoustic and the rest of the 9 phonon modes being optical. Phonon spectra of Pt$_2$CoSn is the most unique among four L2$_1$ phase alloys studied in this work. Six high frequency optical modes originate in the frequency range of 3-5 THz. The rest of the 3 optical phonon modes and 3 acoustic phonon modes are completely imaginary and have negative frequencies. As our hypothesis suggests, the lighter atoms of Y and Z contribute more towards the high frequency phonon modes. The heavier Pt atoms in X sites dominate the imaginary frequencies of optical and acoustic phonon modes with a little contribution from Sn vibrations. This indicates that the vibrations of Pt atoms in the crystal structure are extremely unsustainable and introduce major thermodynamic instability to the crystal structure. Figure 31 shows the thermal properties (G, S and $C_v$) of $Pt_2CoSn$ as a function of temperature in the temperature range of 0-1000 K. As the phonon spectra of $Pt_2CoSn$ consists of imaginary phonon modes, the calculated thermal properties are not truly representative of $Pt_2CoSn$ alloy.

## 3.4 Electrical Conductivity

The electrical conductivity of a material is a property that depends on complex electron transport phenomenon (described by Boltzmann Transport Equation [9] in this work) occurring in the presence of unidirectional electric field, resulting in the drift motion of charge carriers (electrons and holes) in the conduction band of the material [9]. This transport phenomenon of charge carrier relies on multitude of factors such as electronic structure of the crystalline materials, and phonons.

In current work, electron density at $E_F$ is hypothesized to be one of the primary factors affecting electrical conduction in metallic materials. According the observations, the higher the electron density at $E_F$, the higher the electrical conductivity of the material. In half-metallic alloys, there exists a bandgap in the minority spin configuration of electron structure at $E_F$. It reduces the electron density at $E_F$ and adversely affects the electrical conductivity of the materials. The alloys belonging to Heusler alloys have been observed to have half-metallic characteristics. Hence it is expected that the electrical conductivity of the Heusler alloys with bandgap would be lower compared to the ones showing no signs of half-metallicity.

Electron-Phonon (e-ph) interaction is also an important parameter affecting the electrical conductivity of the materials. As the material possesses higher phonon frequencies, the e-ph scattering increases, which reduces the electrical conductivity due to increased resistance to the electron transport.

In addition to the abovementioned parameters, in a practical lattice, there are multitude of lattice imperfections, introducing resistance to the charge carrier transport. These imperfections, being random and circumstantial, are extremely challenging to incorporate into the theoretical simulations and ab initio calculations. Hence, for the sake of simplicity, this work assumes that the alloys are made of an ideal single crystal. The conductivities are calculated with unit relaxation time approximation hence electrical conductivity has a unit of $\frac{1}{\Omega m s}$ instead of $\frac{1}{\Omega m}$. Electrical conductivities are calculated for the temperature range of 10 K to 1000 K.

### 3.4.1 Electrical Conductivities of Half-Heusler Alloys

Figure 32 compares the $\sigma / \tau_0$ for Half-Heusler alloys. Comparison of electrical conductivities of AuCrSn, AuMnGe and PdMnSn shows that the AuMnGe has evidently low electrical conductivity compared to the other two alloys even at temperatures closer to absolute zero. In the analysis of electronic structure of C1$_b$ alloys, alloy AuMnGe has a minority-spin bandgap at E$_F$. AuCrSn and PdMnSn do not have such a bandgap. As observed, the presence of the bandgap in AuMnGe reduces the electron concentration at E$_F$ and causes low electrical conductivity. By comparing the total density of states at E$_F$ for AuCrSn and PdMnSn, it can be observed that AuCrSn has higher states compared to PdMnSn and thus resulting in higher electrical conductivity. Electrical conductivity and electronic thermal conductivities at 300K are given in the table below.

| Alloy | Electrical Conductivity $\left(\frac{1}{\Omega ms}\right)$ | Thermal Conductivity $\left(\frac{W}{m\,K\,s}\right)$ |
|-------|-------------------------------------|--------------------------------------|
| AuCrSn | $1.512 \times 10^{20}$ | $1.143 \times 10^{15}$ |
| AuMnGe | $3.542 \times 10^{19}$ | $3.048 \times 10^{14}$ |
| PdMnSn | $8.571 \times 10^{19}$ | $5.991 \times 10^{14}$ |

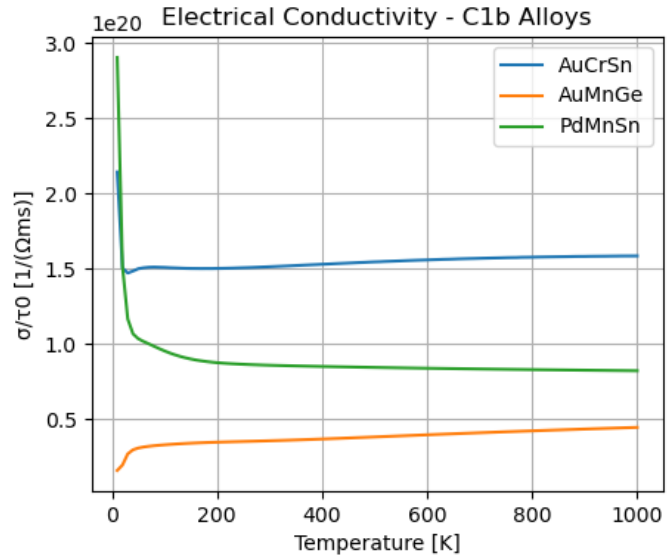

*Figure 32 Electrical Conductivity vs Temperature - Half Heusler Alloys*

The phonon modes of these three alloys play an important role in electrical conduction. It can be observed in the phonon spectra of AuMnGe in Figure 26, that the Mn atoms have phonon modes of frequencies as high as 5-6 THz compared to that of AuCrSn at 4-5 THz. With higher phonon frequencies, e-ph scattering is hypothesized to be higher, increasing the electrical resistance. Due to the higher e-ph interaction and a minority bandgap at $E_F$, AuMnGe was found to have least electrical conductivity among $C1_b$ alloys studied in this work. With the rise in temperature, due to increased kinetic energy of the electrons and atoms, scattering increases substantially. It causes the electrical conductivity to reduce. The same is evident from the electrical conductivity vs temperature plot in Figure 32.

### 3.4.2 Electrical Conductivities of Full-Heusler Alloys

The electrical conductivity of $Au_2MnSn$, $Cu_2NiGe$, $Pd_2NiGe$, and $Pt_2CoSn$ are compared in Figure 33. It can be observed that $Au_2MnSn$ has the highest electrical conductivity at temperatures closer to absolute zero. The electrical conductivity of $Cu_2NiGe$ rapidly increases with the rise in temperature up to 100K and then proceeds to reduce following the common trend observed in Figure 33. $Pt_2CoSn$ was found to have the least electrical conductivity of all four $L2_1$ alloys studied in this work.

None of the four $L2_1$ alloys were found to have bandgap in the minority configuration of their electron structures, and it was concluded in the discussion of electronic structure of these alloys that they do not exhibit half-metallic characteristics.$Cu_2NiGe$ and $Pd_2NiGe$ have same atoms in the Y and Z sites. Despite these similarities, the electrical conductivity of $Cu_2NiGe$ is much higher compared to $Pd_2NiGe$ because of the presence of Cu atoms in $X_1$ and $X_2$ sites. Table 4 shows electrical and thermal conductivities at 300K.

*Table 4 Electrical and Thermal Conductivities of Full Heusler Alloys at 300K*

| Alloy | Electrical Conductivity $\left(\frac{1}{\Omega m s}\right)$ | Thermal Conductivity $\left(\frac{W}{m\,K\,s}\right)$ |
|---|---|---|
| $Au_2MnSn$ | $9.373 \times 10^{19}$ | $6.598 \times 10^{14}$ |
| $Cu_2NiGe$ | $1.347 \times 10^{20}$ | $8.769 \times 10^{14}$ |
| $Pd_2NiGe$ | $8.148 \times 10^{19}$ | $4.982 \times 10^{14}$ |
| $Pt_2CoSn$ | $5.612 \times 10^{19}$ | $4.119 \times 10^{14}$ |

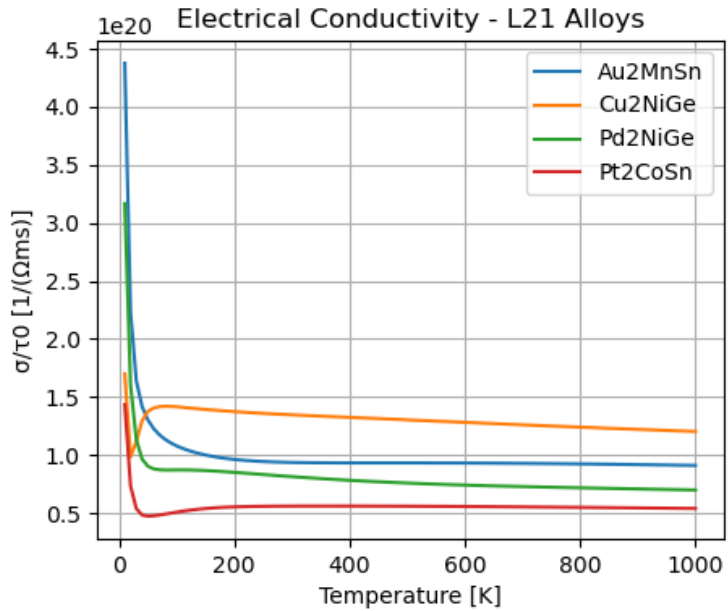

*Figure 33 Electrical Conductivity vs Temperature - Full Heusler Alloys*

In $Au_2MnSn$, s bands of Au and Sn, p bands of Sn and minority d bands of Mn contribute to the conduction bands, which leads to higher electrical conductivity. An insightful picture can be drawn from the phonon modes of these alloys. In $Au_2MnSn$, the heavy Au atom's vibrations dominate the low frequency modes, and the highest frequency optical modes originate at mere 4 THz. The high frequency modes in $Pt_2CoSn$ have frequencies between 4-5 THz. However, the heavy Pt atoms of $X_1$ and $X_2$ sites produce phonon modes of imaginary vibrational frequencies, which introduces thermodynamic instability to the L2$_1$ phase. As the negative modes suggest the existence of vibrations of increasing amplitudes, the scattering events increase and reduce the electrical conductivity.

## 3.5 Thermal Conductivity

This study touches upon only the electronic component of thermal conductivity and not the lattice thermal conductivity and it is calculated with unit relaxation time assumption. Thermal conductivity is a measure of kinetic energy transport in the lattice, which is primarily dependent upon electron-phonon scattering and phonon-phonon scattering [9,22].

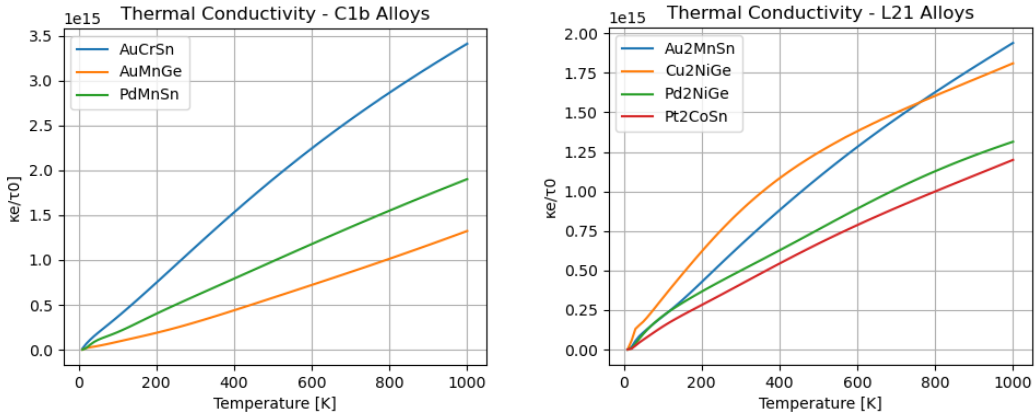

*Figure 34 Electronic Thermal Conductivity vs Temperature*

Electronic Thermal conductivity ($\kappa_e$) is dependent on the electron-phonon scattering. With the rise in temperature, e-ph scattering increases [21]. Higher e-ph scattering allows more energy transfer. Thus, electronic thermal conductivity consistently increases with the temperature. The thermal conductivity vs temperature plots for Half-Heusler and Full-Heusler alloys indicate the same in Figure 34. Electron density in the conduction band is crucial for thermal conduction through e-ph interaction. It is evident that AuMnGe has the lowest $\kappa_e$ due to its half-metallic characteristics and all the alloys follow the same order as electrical conductivity as both entities are primarily reliant on the electron structure.

CHAPTER 4

CONCLUSION

In this work, I calculated the relaxed structures, partial and total density of states, band structures, phonon modes, thermal properties (entropy, specific heat at constant volume and Gibbs free energy), electrical conductivities, and electronic thermal conductivities of the seven alloys AuCrSn, AuMnGe, PdMnSn, $Au_2MnSn$, $Cu_2NiGe$, $Pd_2NiGe$ and $Pt_2CoSn$ of Heusler family. In the electronic structure analysis, I discussed the origin of bandgap in both categories of alloys. The bandgap originated from the d-d hybridization between X-Y atoms in Half-Heusler alloys and between X-X atoms in Full-Heusler alloys. I concluded that AuMnGe was the only alloy among seven alloys to have a bandgap in minority configuration of band structure and was found to be half-metallic. The other six alloys showcased metallic characteristics due to no d-d hybridization. In the analysis of phonon modes, it was found that AuMnGe from $C1_b$ structures and all four alloys from $L2_1$ had unsustainable vibrations from imaginary phonon modes, ensuing phase instability. Electrical conductivities and electronic thermal conductivities of the alloys were primarily reliant of the electronic structure of the alloys and half-metallic system was found to have least electrical conductivity and electronic thermal conductivity. The conductivities of constituent materials also played a role as can be seen in the case of $Cu_2NiGe$ and $Au_2MnSn$.

For the seven alloys considered in this study, it was found that AuMnGe is the only alloy with a minority-spin bandgap. From the assessment of electron structure, it can be concluded that the choice of X and Y atoms should be such that they have more unoccupied minority d bands in valence shell and the valence d band lies in the similar energy range in their density of states plots. From the observations made from this work, it would be safe to conclude that in order to ensure the stability of the composition, it should be taken into consideration that the constituent atoms are not as heavy as Au, Pd and Pt as their vibrations may cause thermodynamic instability to the $C1_b$ or $L2_1$ phase.

The thermal properties – Entropy (S), Specific Heat at constant volume ($C_v$) and Gibbs Free Energy (G) were calculated over the temperature range of 0-1000 K based on the phonon modes of these materials. The obtained results for these properties of AuCrSn and PdMnSn accurately describe the respective materials. Due to the existence of imaginary phonon modes for the lattice of rest of the five materials, the calculated properties, Gibbs Free Energy, Specific heat, and Entropy are not truly representative of the respective alloys. However, the calculated values of G, S and $C_v$ do provide a foundation for further research to be carried out in this area.

REFERENCES

1. " Electronic structure and Slater-Pauling Behaviour in Half-Metallic Heusler Alloys Calculated from First Principles" *Galanakis et al* 2006 *J. Phys. D: Appl. Phys.* **39** 765 https://doi.org/10.1088/0022-3727/39/5/S01

2. Heusler compound. (2023, September 28). In *Wikipedia*. https://en.wikipedia.org/wiki/Heusler_compound

3. Nihat Arıkan, Yıldız, Y.G. & Yıldız, G.D. First-Principles Study on PdMnSn and PtMnSn Compounds in C1b Structure. J. Exp. Theor. Phys. 130, 673–680 (2020). https://doi.org/10.1134/S1063776120050015

4. Matsumoto, A., Kobayashi, K., Ozaki, K., & Nishio, T. (2005). PREPARATION OF Fe2VAl COMPOUND USING MA-PCS PROCESS AND THE THERMOELECTRIC PROPERTY. *Novel Materials Processing by Advanced Electromagnetic Energy Sources*, 377-380. https://doi.org/10.1016/B978-008044504-5/50077-5

5. Giustino, F. (2014). *Materials Modelling using Density Functional Theory*. Oxford University Press.

6. "Electrical Conductivity from First Principles", Zhenkun Yuan (2022).

7. (2007). ELECTRONIC MOTION: DENSITY FUNCTIONAL THEORY (DFT). *Ideas of Quantum Chemistry*, 567-614. https://doi.org/10.1016/B978-044452227-6/50012-0

8. Attarakih, M., Hasseine, A., & Bart, H. (2017). On the Solution of the PBE by Orthogonal Expansion of the Maximum Entropy Functional. *Computer Aided Chemical Engineering*, *40*, 2053-2058. https://doi.org/10.1016/B978-0-444-63965-3.50344-5

9. Wesche, R. (2017). Springer Handbook of Electronic and Photonic Materials. In *Springer handbooks*. https://doi.org/10.1007/978-3-319-48933-9

10. Giannozzi, Paolo & Baroni, Stefano. (2005). Density-Functional Perturbation Theory. 10.1007/978-1-4020-3286-8_11.

11. G. Kresse and J. Hafner, Phys. Rev. B **47**, 558 (1993); ibid. 49, 14251 (1994).

12. G. Kresse and J. Furthmüller, Comput. Mat. Sci. **6**, 15 (1996).

13. G. Kresse and J. Furthmüller, Phys. Rev. B **54**, 11169 (1996).

14. G. Kresse and D. Joubert, Phys. Rev. **59**, 1758 (1999).

15. Togo, A., & Tanaka, I. (2015). First principles phonon calculations in materials science. *Scripta Materialia*, *108*, 1-5. https://doi.org/10.1016/j.scriptamat.2015.07.021

16. V. Wang, N. Xu, J.C. Liu, G. Tang, W.T. Geng, VASPKIT: A User-Friendly Interface Facilitating High-Throughput Computing and Analysis Using VASP Code, **Computer Physics Communications** 267, 108033 (2021). https://doi.org/10.1016/j.cpc.2021.108033

17. "Implementation strategies in phonopy and phono3py", Atsushi Togo, Laurent Chaput, Terumasa Tadano, and Isao Tanaka, J. Phys. Condens. Matter 35, 353001-1-22 (2023). https://dx.doi.org/10.1088/1361-648X/acd831 (Open access)

18. "First-principles Phonon Calculations with Phonopy and Phono3py", Atsushi Togo, J. Phys. Soc. Jpn., 92, 012001-1-21 (2023) https://doi.org/10.7566/JPSJ.92.012001 (Open access)

19. Madsen, G. K., & Singh, D. J. (2006). BoltzTraP. A code for calculating band-structure dependent quantities. *Computer Physics Communications*, *175*(1), 67-71. https://doi.org/10.1016/j.cpc.2006.03.007

20. Madsen, G. K., Carrete, J., & Verstraete, M. J. (2018). BoltzTraP2, a program for interpolating band structures and calculating semi-classical transport coefficients. *Computer Physics Communications*, *231*, 140-145. https://doi.org/10.1016/j.cpc.2018.05.010

21. Band, Y. B., & Avishai, Y. (2013). Spin. *Quantum Mechanics With Applications to Nanotechnology and Information Science*, 159-192. https://doi.org/10.1016/B978-0-444-53786-7.00004-6

22. Chen, L., Chen, S., & Hou, Y. (2019). Understanding the thermal conductivity of Diamond/Copper composites by first-principles calculations. *Carbon*, *148*, 249-257. https://doi.org/10.1016/j.carbon.2019.03.051